\documentclass[preprint2]{aastex63}

\def \Physics {Department of Physics, University of Michigan, Ann Arbor, MI 48109, USA}
\def \Astronomy {Department of Astronomy, University of Michigan, Ann Arbor, MI 48109, USA}
\def \caltech {Division of Geological and Planetary Sciences, California Institute of Technology, Pasadena, CA 91125}

\newcommand\slope{\gamma} 

\received{}
\revised{}
\accepted{}
\submitjournal{ApJ Letters}

\shorttitle{C2 detections of 2I}
\shortauthors{Lin et al.}

\graphicspath{{./}{figures/}}

\begin{document}

\title{Detection of Diatomic Carbon in 2I/Borisov}

\correspondingauthor{Hsing Wen Lin}
\email{hsingwel@umich.edu}

\author{Hsing~Wen~Lin}
\affiliation{\Physics}

\author{Chien-Hsiu Lee}
\affiliation{NSF National Optical Infrared Astrophysical Research Laboratory, Tucson, AZ 85719, USA}

\author{D.~W.~Gerdes}
\affiliation{\Physics}
\affiliation{\Astronomy}

\author{Fred C. Adams}
\affiliation{\Physics}
\affiliation{\Astronomy}

\author{Juliette Becker}
\affiliation{\caltech}

\author{Kevin Napier}
\affiliation{\Physics}

\author{Larissa Markwardt}
\affiliation{\Astronomy}

\begin{abstract}

2I/Borisov is the first-ever observed interstellar comet (and the second detected interstellar object). It was discovered on 30 August 2019 and has a heliocentric orbital eccentricity of $\sim 3.35$, corresponding to a hyperbolic orbit that is unbound to the Sun. Given that it is an interstellar object, it is of interest to compare its properties -- such as composition and activity -- with the comets in our Solar System. This study reports low-resolution optical spectra of 2I/Borisov. The spectra were obtained by the MDM observatory Hiltner 2.4m telescope/Ohio State Multi-Object Spectrograph (on October 31.5 and November 4.5, 2019 UT). The wavelength coverage spanned from $3700\AA$ to $9200\AA$. The dust continuum reflectance spectra of 2I/Borisov show that the spectral slope is steeper in the blue end of the spectrum (compared to the red). The spectra of 2I/Borisov clearly show CN emission at $3880\AA$, as well as C$_2$ emission at both $4750\AA$ and $5150\AA$. Using a Haser model to covert the observed fluxes into estimates for the molecular production rates, we find $Q$(CN) = $2.4 \pm 0.2 \times 10^{24}$~s$^{-1}$, and $Q$(C$_2$) = $5.5 \pm 0.4 \times 10^{23}$~s$^{-1}$ at the heliocentric distance of 2.145 au. Our $Q$(CN) estimate is consistent with contemporaneous observations, and the $Q$(C$_2$) estimate is generally below the upper limits of previous studies. We derived the ratio $Q$(C$_2$)/$Q$(CN) = $0.2 \pm 0.1$, which indicates that 2I/Borisov is depleted in carbon chain species, but is not empty. This feature is not rare for the comets in our Solar System, especially in the class of Jupiter Family Comets. 

\end{abstract}



\section{Introduction} \label{sec:intro}

The detection of the first interstellar object (ISO), known as 'Oumuamua \citep{Meech17}, has ushered in a new field of astronomy -- the search and study of interloping minor bodies passing through the Solar System. Within two years of the discovery of the first ISO, the second ISO, known as 2I/Borisov \citep{Guzik19}, was discovered. 2I/Borisov is markedly different from 'Oumuamua. The new object is likely larger and displays a prominent cometary tail \citep{Bolin19, Jewitt19, Jewitt19b, Lee19, Ye19}, and thus provides a unique opportunity to study the composition of comets originating from other planetary systems. 

Upon the arrival of 2I/Borisov, it was studied observationally using  follow-up imaging and spectroscopy. \citet{deLeon19} obtained optical spectra with the 10-m GTC telescope and reported a spectral shape similar to that of D-type asteroids. \citet{fitz19} acquired optical spectra with the William Herschel Telescope (specifically at wavelengths shorter than the observations of \citealt{deLeon19}) and presented the first detection of CN emission at $3880\AA$. \citet{Kareta19} carried out optical spectroscopic observations of the comet with MMT and LBT at the heliocentric distance above 2.4~au but only obtained non-detection of diatomic Carbon C$_2$. From the upper limit of $Q$(CN)/$Q$(C$_2$) ratio, \citet{Kareta19} concluded that 2I/Borisov is a carbon-depleted comet, similar to the Jupiter family comets that are seen in our Solar System. \citet{Opitom19} investigated the C$_2$ emission from the WHT observations when 2I/Borisov was 2.36~au away from the Sun, but also did not detect C$_2$. Nevertheless, both \citet{Kareta19} and \citet{Opitom19} concluded that 2I/Borisov is a carbon-depleted comet, and possibly exhausted its surface material before leaving its natal planetary system. As 2I/Borisov is still on an inbound trajectory toward the Sun, its cometary activity will increase, and we are likely to detect C$_2$ emission as the object approaches perihelion. In addition to CN and C$_2$, [OI] at 6300\AA~was also detected and indicates that 2I/Borisov may also contain water ice \citep{McKay19}. However, the absorption features of water ice in near-infrared have not been detected on the observations before early October \citep{Yang19}. 

This present work has three principal objectives. The first is to conduct continuous monitoring of 2I/Borisov, with the goal of revealing C$_2$ emission. The second is to use the observed $Q$(CN)/$Q$(C2) ratio to constrain the surface properties of 2I/Borisov. The final objective is to investigate the evolution of both production rates $Q$(CN) and $Q$(C2) as a function of heliocentric distance, specifically as 2I/Borisov approaches perihelion.

\section{Observation and Data Reduction} \label{sec:obs}

2I/Borisov was observed with the Hiltner 2.4m telescope and the Ohio State Multi-Object Spectrograph (OSMOS) \citep{Martini11} on 2019 October~31.5 and November~4.5 UT via the queue observations of MDM observatory. The observational circumstances are listed in Table~\ref{tab1}. The high throughput triplet prism mode of OSMOS was used to obtain low-resolution spectra. The triplet prism produces variable resolution of $\lambda/\Delta\lambda = 400 - 60$ across the wavelength range from 3600 to 10000\AA~\citep{Martini11}. The higher resolution of the triplet prism for the short wavelengths is ideal for the detection of gas emissions, i.e., CN and C$_2$, and the wide wavelength coverage is suitable to study the spectral type of the dust coma. The Hiltner 2.4m telescope cannot guide non-sidereally. As a result, we observed with sidereal tracking and aligned the slit with the motion direction of 2I/Borisov (at PA = 143 degrees). On 2019 October~31 UT we obtained six exposures and November~4 UT eleven exposures. Each exposure were 300-seconds and obtained with a 3'' wide slit.    

Along with 2I/Borisov, we also observed G191-B2B (October~31 UT) and BD28-4211 (November~4 UT) as flux standards, planetary nebula IC351 as a wavelength calibrator, as well as G2 stars HIP117367 (October~31 UT) and HIP117537 (November~4 UT) as Solar analogs.

\begin{figure*}
\includegraphics[width = 1\textwidth]{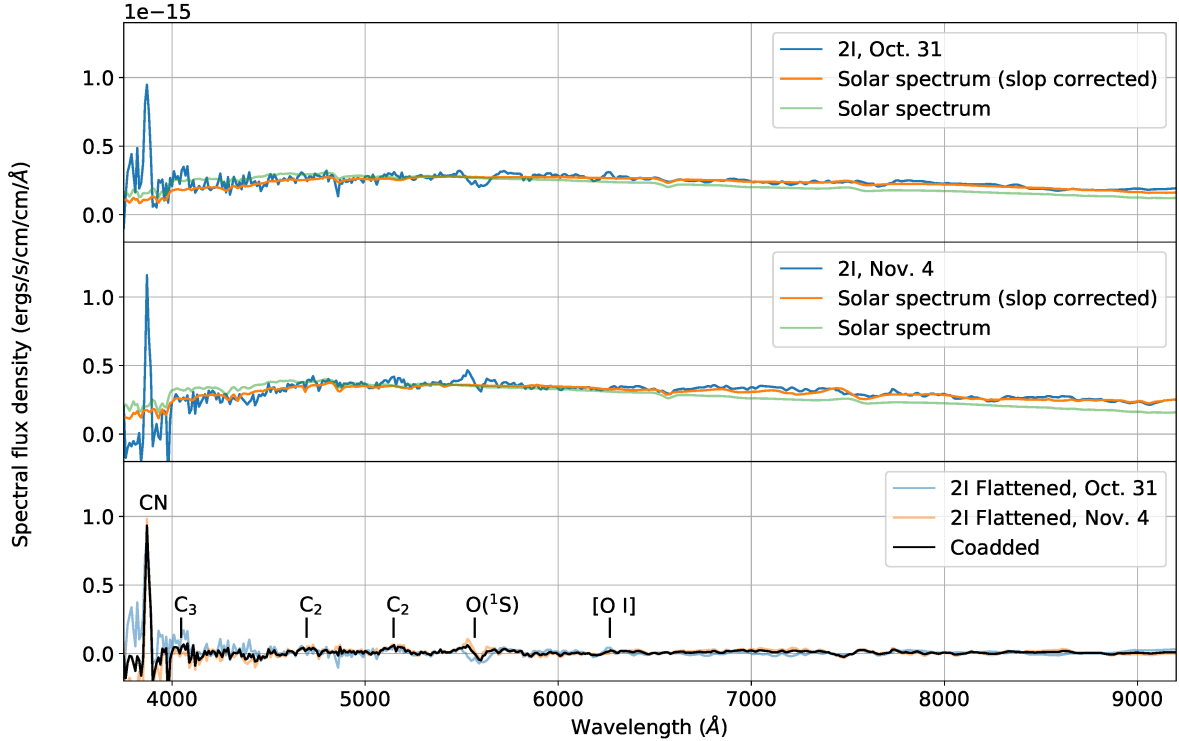}
\caption{The 2I/Borisov spectra obtained by the Hiltner 2.4m telescope/OSMOS on \textbf{Top}: October~31 and \textbf{Middle}: November~4, 2019. The fitted solar analog spectra and slope corrected (see Section~\ref{sec:dust}) solar spectra are also shown.  The \textbf{Bottom} panel shows the flatten spectra with Solar spectrum removed. Noted that the absorption/emission feature around 5600\AA~is the O($^1$S) band due to the non-proper sky subtraction, not the feature of the comet.
\label{fig1}}
\end{figure*}

Since the comet was moving across the slit, we were able to reconstruct the local sky background by combining the set of exposures. We subtracted bias and sky, and then extracted one-dimensional spectra with 27.8'' widths to maximize the signal of the comet. The one-dimensional spectra had wavelength solutions derived via comparing the emission lines of planetary nebula IC351 \citep{Feibelman96} and also flux calibrated with a standard KPNO extinction correction applied. Although the OSMOS/triplet prism has a wavelength range from 3600 to 10000\AA, we were only able to identify the emission lines between 3835\AA~(H9) and 9069\AA~([SIII]). Therefore we chopped the one-dimensional spectra, and only keep the wavelength between 3750\AA~and 9200\AA~to avoid too much extrapolation. The remaining 3750\AA~to 9200\AA~range should have sufficient accurate wavelength solution. The final reduction results of 2I/Borisov spectra are shown in Figure~\ref{fig1}.

\begin{table*}
\caption{Log of observations and production rates of 2I/Borisov \label{tab1}}
\begin{tabular*}{\textwidth}{l @{\extracolsep{\fill}}lccccccc}
\hline
Date (UT) & r$_h$ (au) $^a$ & $\Delta$ (au) $^b$ & Exp Time (s) & Airmass & $Q$(CN) (s$^{-1}$)& $Q$(C$_2$) (s$^{-1}$) $^c$ \\
\hline
\hline
2019 Oct. 31.5 & 2.177 & 2.437 & $6 \times 300$ & 1.64-1.46 &$(2.0 \pm 0.2) \times 10^{24}$ & -- \\
2019 Nov. 04.5 & 2.145 & 2.373 & $11 \times 300$ & 1.91-1.47 &$(2.4 \pm 0.2) \times 10^{24}$ & $(5.5 \pm 0.4) \times 10^{23}$ \\

\hline
\end{tabular*}
$^a$ Heliocentric distance\\
$^b$ Geocentric distance\\
$^c$ $Q$(C$_2$) estimated from co-added all of the usable spectra.
\end{table*}

\section{Analysis and Results} \label{sec:res}

\subsection{The Dust Continuum} \label{sec:dust}

\begin{figure}
\includegraphics[width = .5\textwidth]{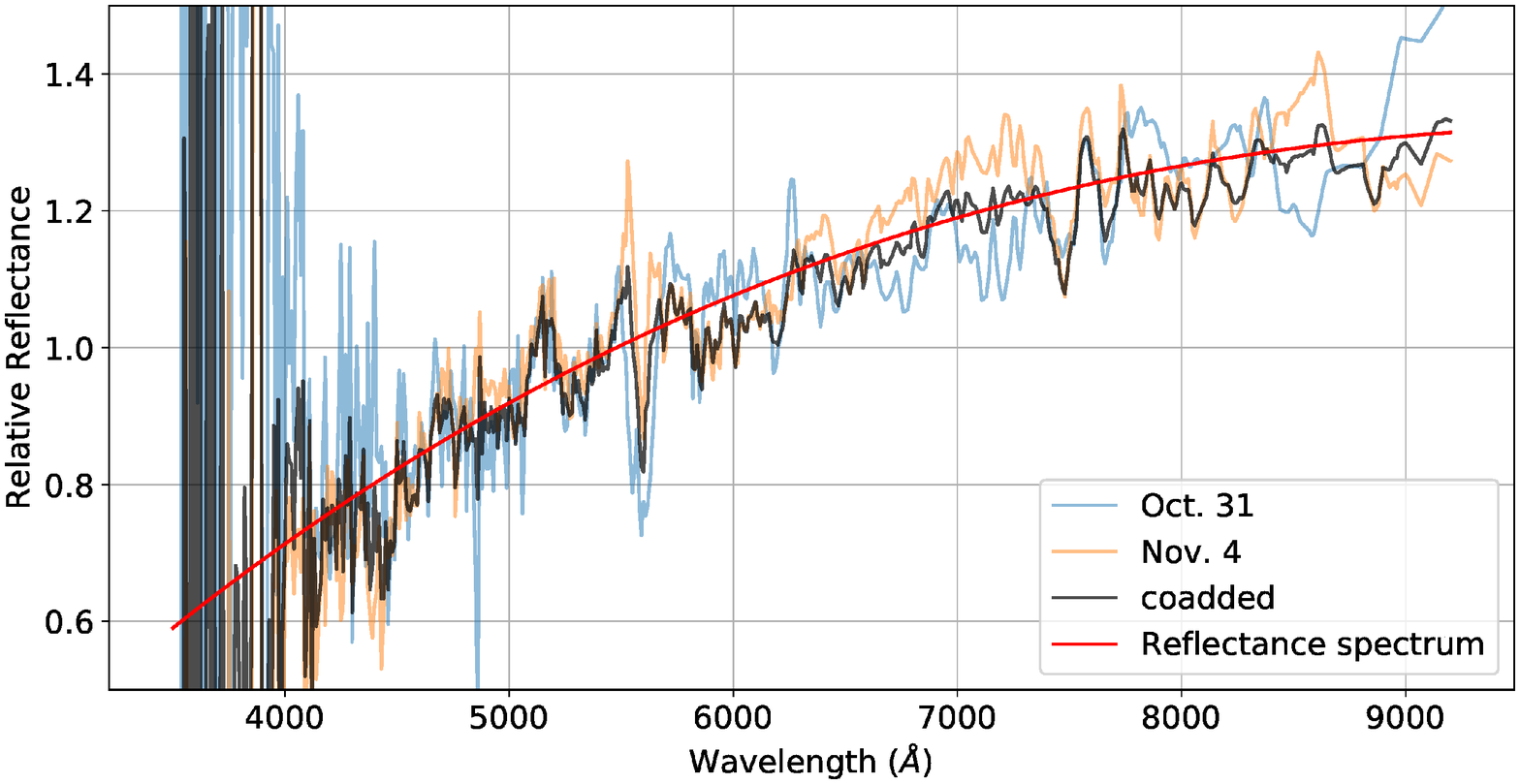}
\caption{The reflectance spectrum of 2I/Borisov. A cubic spline is also plotted to model the trends of spectral slope variation.
\label{fig2}}
\end{figure}

The low-resolution spectra from 3750\AA~to 9200\AA~allows the measurement of the coma dust reflectance across the whole visible wavelength range. We divided the 2I/Borisov spectra (Figure~\ref{fig1}, 2I) by solar analogs (Figure~\ref{fig1}, Solar spectrum) in order to normalize the reflectance spectrum, where the equality point is taken to be 5500\AA. We then co-added the spectra observed on October~31 and November~4 by weighting them by their total exposure time. The general slope of the reflectance spectrum varies with wavelength. As a result, we fitted the spectrum with a cubic spline to specify the spectral slope as a function of wavelength. The result is shown in Figure~\ref{fig2}. The spectral slope is steeper in the shorter wavelength range than in the longer wavelength range. More specifically, found an average slope of $19.3\%/10^3 \AA$ in the range of $3900\AA~< \lambda < 6000\AA$, and an average slope of $9.2\%/10^3 \AA$ in the range of $5500\AA~< \lambda < 9000\AA$. This result is consistent with both the blue-end slope reported by \citet{fitz19} ($19.9 \pm 1.5\%/10^3 \AA$), \citet{Kareta19} ($22\%/10^3 \AA$) and red end-slope reported by \citet{deLeon19} ($10 \pm 1\%/10^3 \AA$) and by \citet{Kareta19} ($11\%/10^3 \AA$). Such a result agrees with the spectral behavior of scattered light of micron-sized coma dust grains, which are commonly found within normal solar system comets \citep{Jewitt86}.

\subsection{CN Emission and Production Rate} \label{sec:cn}

The 2I/Borisov spectra can be flattened by subtracting the solar spectrum, with the correction applied for the spectral slope, as described in Section~\ref{sec:dust}. The flattened spectra are shown in the bottom panel of Figure~\ref{fig1}. We identified gas emission lines in these processed spectra. The first feature is the clear CN(0-0) emission at 3880\AA. The flux of CN emission was directly measured by fitting a Gaussian function and giving a value of $(2.3 \pm 0.2) \times 10^{-14}$ ergs~s$^{-1}$~cm$^{-2}$ on October~31 and $(2.9 \pm 0.2) \times 10^{-14}$ ergs~s$^{-1}$~cm$^{-2}$ on November~4.

To convert the observed flux into a gas production rate, we first obtained fluorescence efficiency of CN from \citet{Schleicher10} to calculate the number of CN molecules within the extraction aperture. We then used a simple Haser model \citep{Haser57} from \tt{sbpy} \normalfont \citep{Mommert19} to calculate production rate of CN. The scale lengths of the parent (HCN, assuming CN gas is only generated by the photodissociation of this parent molecule) and daughter molecule (CN) were taken from \citet{AHearn95}, and we used the outflow velocity $0.85 \times r_h^{-0.5} = 0.58$ km~s$^{-1}$ \citep{Cochran93}. Because the Haser model is spherically symmetric, we integrate the Haser model with a 13.9'' radius circular aperture and adjust the value to the equivalent area of our 3'' by 27.8'' aperture to derive the corresponding production rate of CN. We found $Q$(CN) $= (2.0 \pm 0.2)  \times 10^{24}$ s$^{-1}$ on October~31 and $(2.4 \pm 0.2)  \times 10^{24}$ s$^{-1}$ on November~4. This result is consistent with the previous $Q$(CN) estimates from October 2019 \citep{Kareta19, Opitom19} and suggests that the CN production rate of 2I/Borisov did not increase dramatically as it approaches perihelion (over this time interval). 

\subsection{Possible C$_3$ Emission} \label{sec:c3}

\begin{figure}
\includegraphics[width = .5\textwidth]{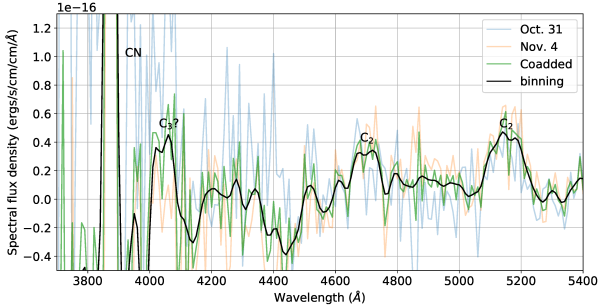}
\caption{The magnified spectra of 2I/Borisov from 3750\AA~to 5400\AA. 
\label{fig3}}
\end{figure}

In addition to the CN emission, we used the co-added spectrum to search for other weaker features across the entire wavelength range. We found the excess emission near the wavelengths corresponding to C$_3$ and C$_2$ bands. The magnified (zoomed-in) spectrum in the wavelength range 3750\AA~to 5400\AA~is shown in Figure~\ref{fig3}. The excess at 4050\AA~matches the wavelength appropriate for C$_3$ emission. However, considering that this excess could be driven by the noisy data taken on October~31 (see Figure~\ref{fig3}), we do not have sufficient confidence to claim the detection of C$_3$. However, if the possible C$_3$ emission is real, we estimate a flux $F = (2 \pm 1) \times 10^{-15}$ ergs~s$^{-1}$~cm$^{-2}$ (extracted from the co-added but non-binned spectrum).

We calculated the C$_3$ production rate using the same method of Section~\ref{sec:cn} with fluorescence efficiency of C$_3$ and the scale lengths of the parent/daughter molecule adopted from \citet{AHearn95}. Assuming that the outflow velocity is 0.58 km~s$^{-1}$ and that (again) this emission is real, we find  $Q$(C$_3$) $= (3 \pm 1) \times 10^{22}$ s$^{-1}$. Note that this value is below the upper limit of $< 2 \times 10^{23}$ s$^{-1}$ found in \citet{Opitom19}, and hence consistent. 

\subsection{C$_2$ Emission and Production Rate} \label{sec:c2}

The other two excesses are located at 4700\AA~and 5150\AA, which are matched the locations of C$_2(\Delta v = 1)$ and C$_2(\Delta v = 0)$ emissions, respectively. Unlike the excess near C$_3$ band, however, the C$_2$ emission excesses were observed in both spectra; this consistency indicates that these features represent the detection of C$_2$ emission. The only potential issue is that the shapes of C$_2$ emission features are not exactly what is expected, in that we do not see sharp breaks in the longer wavelength sides. However, considering the effects of binning data, low spectral resolution, and low signal-to-noise ratio, it is highly possible to lose or smooth out the shapes of the C$_2$ emission features. 

We measured the C$_2$ emission fluxes via the co-added but unbinned spectrum, and derived the C$_2(\Delta v = 0)$ flux $F= (5.7 \pm 0.4) \times 10^{-15}$ ergs~s$^{-1}$~cm$^{-2}$ and C$_2(\Delta v = 1)$ flux $F= (3.0 \pm 0.6) \times 10^{-15}$ ergs~s$^{-1}$~cm$^{-2}$. The flux ratio between C$_2(\Delta v = 0)$ and C$_2(\Delta v = 1)$ is about 1.9. This result is  consistent with the fluorescence efficiency L/N of C$_2(\Delta v = 0) = 4.5 \times 10^{-23}$ erg~s$^{-1}$ and L/N of C$_2(\Delta v = 1) = 2.4 \times 10^{-23}$ erg~s$^{-1}$ \citep{deAlmeida89}, which indicates that the C$_2(\Delta v = 0)$ emission should be about 1.9 times stronger than the emission of C$_2(\Delta v = 1)$. We consider this agreement to be additional evidence suggesting that our C$_2$ detections are real.

We calculated the C$_2$ production rate via C$_2(\Delta v = 0)$ emission with fluorescence efficiency of C$_2$ and the scale lengths of the parent/daughter molecule adopted from \citet{AHearn95}. Assuming that the outflow velocity is 0.58 km~s$^{-1}$, we obtain the production rate $Q$(C$_2$) $= (5.5 \pm 0.4) \times 10^{23}$ s$^{-1}$.  This result is below the upper limit estimated by \citet{Opitom19}, and also below or consistent with the measurements of \citet{Kareta19}. The only discrepancy is the LBT observation on Oct. 10, which is substantially below our estimated value. However, 2I/Borisov was closer to the Sun by 0.245~au during this new measurement (compared to Oct. 10), so that the C$_2$ production rate could have increased enough to be detectable.

\subsection{Searching for Additional Spectral Features} \label{sec:OI}
Besides the main carbon species, we also searched other spectral features such as [O I] at 6300\AA~and NH2 emissions across 5500\AA~to 7500\AA. The spectral beyond 5200\AA~is mostly flat, except a weak signal at 6300\AA~in the spectrum taken on October~31, 2019 UT (see Figure~\ref{fig2}). This signal should belong to [O I] emission. However, since our spectral resolution is low at 6300\AA~($\lambda/\Delta\lambda \sim 120$), we can not distinguish the cometary [O I] feature from telluric [O I] emission. Therefore, this [O I] feature is very likely not fully belong to the comet.

\section{Discussion} \label{sec:dis}

In this section we estimate the C$_2$ production rate as a function of the heliocentric distance ($r_h$). \citet{Kareta19} obtained an upper limit of $Q$(C$_2$) $< 1.62 \times 10^{23}$s$^{-1}$, when the comet was at distance $r_h$ = 2.39~au. This study finds $Q$(C$_2$) $= (5.5 \pm 0.1) \times 10^{23}$ s$^{-1}$ when the distance has decreased to $r_h$ = 2.145~au. As a result, the rate $Q$(C$_2$) increased by an increment of at least $3 \times 10^{23}$ s$^{-1}$ as the body traveled a distance of 0.245~au closer to the Sun. 
We can calculate the power-law slope $\slope$ for C$_2$ production rate as a function of $r_h$,  which is given by 
\begin{equation}
\slope = {\log~Q(r_{h_1}) - \log~Q(r_{h_0}) \over 
\log~r_{h_1} - \log~r_{h_0}} \,. \\
\end{equation}
With our lower bound of $Q$(C$_2$) $\approx 5.1 \times 10^{23}$ s$^{-1}$, and the upper limit adopted from \citet{Kareta19}, we find a bound on the slope $\slope < -10.6$. This slope is significantly steeper than those of other comets in our Solar System, such as $\slope = -3.56 \pm 0.16$ for C/2013 R1 (Lovejoy) \citep{Opitom15}, $\slope = -4.10 \pm 0.10$ for 103P/Hartley 2 \citep{Knight13}, and $\slope = -2.60$ to $-4.39$ for 81P/Wild 2 \citep{lin12}. 

The steeper slope for 2I/Borisov suggests that its C$_2$ production rate might be more sensitive to the heliocentric distance $r_h$ than the Solar System comets. On the other hand, if we assume that 2I/Borisov has a $Q$(C$_2$) slope that is similar to Solar System comets (i.e., $\slope=-4$), then the rate $Q$(C$_2$) would be about $3.5 \times 10^{23}$ s$^{-1}$ in the middle of October. This rate is thus below the MMT upper limit measurement of $Q<4.4 \times 10^{23}$ s$^{-1}$ (found on October~9, 2019) but higher than the LBT upper limit. With the limited measurements taken to date, it is not possible to draw a firm conclusion on the slope of C$_2$ production. Follow-up observations and measurements of the C$_2$ production rate will be needed to solidify these results.

In contrast to the case of C$_2$, the production rate of CN has not changed dramatically with decreasing distance $r_h$ (see   Section~\ref{sec:cn}). We adopted all of the $Q$(CN) measurements from \citet{Kareta19} and \citet{Opitom19}, and find a power-law slope $\slope = -2 \pm 1$ for $Q$(CN) as a function of $r_h$. This result for the slope of the production rate $Q$(CN) is similar to that of other Solar System comets, where $\slope=-2.60 \pm 0.17$ for C/2013 R1 (Lovejoy) \citep{Opitom15}, $\slope=-3.34 \pm 0.18$ for 103P/Hartley 2 \citep{Knight13}, and $\slope$ = $-3.68$ to $-2.58$ for 81P/Wild 2 \citep{lin12}.

Since the observations detected both CN and C$_2$ emission, we determine the ratio of rates $Q$(C$_2$)/$Q$(CN) = $0.2 \pm 0.1$. In comparing this value with the observations taken previously, we find that it is below the estimated upper limit of $< 0.3$ by \citet{Opitom19}, but higher than the upper limit of $< 0.095$ from \citet{Kareta19}. Nevertheless, our result of $Q$(C$_2$)/$Q$(CN) $\sim 0.2$ is still consistent with the classification of 2I/Borisov as a carbon-chain depleted comet. This type of object is more commonly found among the group of Jupiter Family Comets (JFCs), in contrast to the long period comets (LPCs) \citep{Cochran12, AHearn95}.

We also note that since the C$_2$ has been detected, the abundances of carbon-chain species are low but not zero. Moreover, all of the current observations of 2I/Borisov, including the reflectance spectrum of its dust coma, the detections of CN and C$_2$ emission, and the $Q$(C$_2$)/$Q$(CN) ratio, indicate that its properties are similar to ordinary comets found in our Solar System. This finding, in turn, suggests that the natal disk of 2I/Borisov formed could have similar chemical composition with our Solar System.

\section{Summary} \label{sec:sam} 
In this study, we report the spectroscopic observations of 2I/Borisov on October~31.5 and November~4.5 UT using the Hiltner 2.4m telescopes and the OSMOS Spectrograph on MDM observatory. We find that the dust coma reflectance is a function of wavelength, which is steeper for shorter wavelengths ($9.2\%/10^3 \AA$) and shallower at longer wavelengths ($19.3\%/10^3 \AA$). Emission from both CN and C$_2$ were detected, with possible but unconfirmed C$_3$ emission. We estimated the CN production rate $Q$(CN) $= (2.0 \pm 0.2)  \times 10^{24}$~s$^{-1}$ on Nov. 1  and $Q$(CN) $=(2.4 \pm 0.2) \times 10^{24}$~s$^{-1}$ on Nov. 5. The C$_2$ production rate was also estimated by co-adding the spectra on October~31 and November~4, with the value of $Q$(C$_2$) $= (5.5 \pm 0.4) \times 10^{23}$~s$^{-1}$. Comparing our production rates with the upper limit obtained by \citet{Kareta19}, we find that the rate $Q$(C$_2$) of 2I/Borisov might be relatively more sensitive to heliocentric distance than other Solar System comets. We computed the ratio $Q$(C$_2$)/$Q$(CN) = $0.2 \pm 0.1$, which indicates that 2I/Borisov is a carbon-chain depleted comet. Given that most of the properties currently known about 2I/Borisov are similar to known Solar System comets, this interstellar visitor is likely to have formed within a planetary system much like our own.

\acknowledgments

We thank Mario Mateo, Christopher Miller, Jules Halpern and Eric Galayda for making the comet observations as scheduled. We thank Paul Martini and John Thorstensen for their advice on OSMOS configurations and observations. We thank Justin Rupert, Ryan Chornock and the observational astronomy class of Ohio University for the training of MDM telescopes operation. We thank Zhong-Yi Lin for comments on our data. This work is based on observations obtained at the MDM Observatory, operated by Dartmouth College, Columbia University, Ohio State University, Ohio University, and the University of Michigan.
This material is based upon work supported by the National Aeronautics and Space Administration under Grant No. NNX17AF21G issued through the SSO Planetary Astronomy Program and by NSF grant AST-1515015.

\facility{MDM:Hiltner (OSMOS)}
\software{Scipy, Astropy, Matplotlib, Jupyter, sbpy}

\bibliographystyle{aasjournal}

\bibliography{sample63}{}

\begin{thebibliography}{}
\expandafter\ifx\csname natexlab\endcsname\relax\def\natexlab#1{#1}\fi
\providecommand{\url}[1]{\href{#1}{#1}}
\providecommand{\dodoi}[1]{doi:~\href{http://doi.org/#1}{\nolinkurl{#1}}}
\providecommand{\doeprint}[1]{\href{http://ascl.net/#1}{\nolinkurl{http://ascl.net/#1}}}
\providecommand{\doarXiv}[1]{\href{https://arxiv.org/abs/#1}{\nolinkurl{https://arxiv.org/abs/#1}}}

\bibitem[{{A'Hearn} {et~al.}(1995){A'Hearn}, {Millis}, {Schleicher}, {Osip}, \&
  {Birch}}]{AHearn95}
{A'Hearn}, M.~F., {Millis}, R.~C., {Schleicher}, D.~O., {Osip}, D.~J., \&
  {Birch}, P.~V. 1995, \icarus, 118, 223, \dodoi{10.1006/icar.1995.1190}

\bibitem[{{Bolin} {et~al.}(2019){Bolin}, {Lisse}, {Kasliwal}, {Quimby}, {Tan},
  {Copperwheat}, {Lin}, {Morbidelli}, {Bauer}, {Burdge}, {Coughlin},
  {Fremling}, {Itoh}, {Koss}, {Masci}, {Maeno}, {Mamajek}, {Marocco}, {Murata},
  {Sitko}, {Stern}, {Walters}, {Yan}, {Andreoni}, {Bhalerao}, {Bodewits}, {De},
  {Deshmukh}, {Bellm}, {Blagorodnova}, {Buzasi}, {Cenko}, {Chang},
  {Chojnowski}, {Dekany}, {Duev}, {Graham}, {Juric}, {Kramer}, {Kulkarni},
  {Kupfer}, {Mahabal}, {Neill}, {Ngeow}, {Penprase}, {Riddle}, {Rodriguez},
  {Rosnet}, {Sollerman}, \& {Soumagnac}}]{Bolin19}
{Bolin}, B.~T., {Lisse}, C.~M., {Kasliwal}, M.~M., {et~al.} 2019, arXiv
  e-prints, arXiv:1910.14004.
\newblock \doarXiv{1910.14004}

\bibitem[{{Cochran} {et~al.}(2012){Cochran}, {Barker}, \& {Gray}}]{Cochran12}
{Cochran}, A.~L., {Barker}, E.~S., \& {Gray}, C.~L. 2012, \icarus, 218, 144,
  \dodoi{10.1016/j.icarus.2011.12.010}

\bibitem[{{Cochran} \& {Schleicher}(1993)}]{Cochran93}
{Cochran}, A.~L., \& {Schleicher}, D.~G. 1993, \icarus, 105, 235,
  \dodoi{10.1006/icar.1993.1121}

\bibitem[{{de Almeida} {et~al.}(1989){de Almeida}, {Singh}, \&
  {Burgoyne}}]{deAlmeida89}
{de Almeida}, A.~A., {Singh}, P.~D., \& {Burgoyne}, C.~M. 1989, Earth Moon and
  Planets, 47, 15, \dodoi{10.1007/BF00056328}

\bibitem[{{de Le{\'o}n} {et~al.}(2019){de Le{\'o}n}, {Licandro},
  {Serra-Ricart}, {Cabrera-Lavers}, {Font Serra}, {Scarpa}, {de la Fuente
  Marcos}, \& {de la Fuente Marcos}}]{deLeon19}
{de Le{\'o}n}, J., {Licandro}, J., {Serra-Ricart}, M., {et~al.} 2019, Research
  Notes of the American Astronomical Society, 3, 131,
  \dodoi{10.3847/2515-5172/ab449c}

\bibitem[{{Feibelman} {et~al.}(1996){Feibelman}, {Hyung}, \&
  {Aller}}]{Feibelman96}
{Feibelman}, W.~A., {Hyung}, S., \& {Aller}, L.~H. 1996, \mnras, 278, 625,
  \dodoi{10.1093/mnras/278.2.625}

\bibitem[{{Fitzsimmons} {et~al.}(2019){Fitzsimmons}, {Hainaut}, {Meech},
  {Jehin}, {Moulane}, {Opitom}, {Yang}, {Keane}, {Kleyna}, {Micheli}, \&
  {Snodgrass}}]{fitz19}
{Fitzsimmons}, A., {Hainaut}, O., {Meech}, K.~J., {et~al.} 2019, \apjl, 885,
  L9, \dodoi{10.3847/2041-8213/ab49fc}

\bibitem[{{Guzik} {et~al.}(2019){Guzik}, {Drahus}, {Rusek}, {Waniak},
  {Cannizzaro}, \& {Pastor-Marazuela}}]{Guzik19}
{Guzik}, P., {Drahus}, M., {Rusek}, K., {et~al.} 2019, Nature Astronomy, 467,
  \dodoi{10.1038/s41550-019-0931-8}

\bibitem[{{Haser}(1957)}]{Haser57}
{Haser}, L. 1957, Bulletin de la Societe Royale des Sciences de Liege, 43, 740

\bibitem[{{Jewitt} {et~al.}(2019){Jewitt}, {Hui}, {Kim}, {Mutchler}, {Weaver},
  \& {Agarwal}}]{Jewitt19b}
{Jewitt}, D., {Hui}, M.-T., {Kim}, Y., {et~al.} 2019, arXiv e-prints,
  arXiv:1912.05422.
\newblock \doarXiv{1912.05422}

\bibitem[{{Jewitt} \& {Luu}(2019)}]{Jewitt19}
{Jewitt}, D., \& {Luu}, J. 2019, \apjl, 886, L29,
  \dodoi{10.3847/2041-8213/ab530b}

\bibitem[{{Jewitt} \& {Meech}(1986)}]{Jewitt86}
{Jewitt}, D., \& {Meech}, K.~J. 1986, \apj, 310, 937, \dodoi{10.1086/164745}

\bibitem[{{Kareta} {et~al.}(2019){Kareta}, {Andrews}, {Noonan}, {Harris},
  {Smith}, {O'Brien}, {Sharkey}, {Reddy}, {Springmann}, {Lejoly}, {Volk},
  {Conrad}, \& {Veillet}}]{Kareta19}
{Kareta}, T., {Andrews}, J., {Noonan}, J.~W., {et~al.} 2019, arXiv e-prints,
  arXiv:1910.03222.
\newblock \doarXiv{1910.03222}

\bibitem[{{Knight} \& {Schleicher}(2013)}]{Knight13}
{Knight}, M.~M., \& {Schleicher}, D.~G. 2013, \icarus, 222, 691,
  \dodoi{10.1016/j.icarus.2012.06.004}

\bibitem[{Lee {et~al.}(2019)Lee, Lin, Chen, \& Yen}]{Lee19}
Lee, C.-H., Lin, H.-W., Chen, Y.-T., \& Yen, S.-F. 2019, Research Notes of the
  {AAS}, 3, 184, \dodoi{10.3847/2515-5172/ab5f69}

\bibitem[{{Lin} {et~al.}(2012){Lin}, {Lara}, {Vincent}, \& {Ip}}]{lin12}
{Lin}, Z.~Y., {Lara}, L.~M., {Vincent}, J.~B., \& {Ip}, W.~H. 2012, \aap, 537,
  A101, \dodoi{10.1051/0004-6361/201116848}

\bibitem[{{Martini} {et~al.}(2011){Martini}, {Stoll}, {Derwent}, {Zhelem},
  {Atwood}, {Gonzalez}, {Mason}, {O'Brien}, {Pappalardo}, {Pogge}, {Ward}, \&
  {Wong}}]{Martini11}
{Martini}, P., {Stoll}, R., {Derwent}, M.~A., {et~al.} 2011, \pasp, 123, 187,
  \dodoi{10.1086/658357}

\bibitem[{{McKay} {et~al.}(2019){McKay}, {Cochran}, {Dello Russo}, \&
  {DiSanti}}]{McKay19}
{McKay}, A.~J., {Cochran}, A.~L., {Dello Russo}, N., \& {DiSanti}, M. 2019,
  arXiv e-prints, arXiv:1910.12785.
\newblock \doarXiv{1910.12785}

\bibitem[{{Meech} {et~al.}(2017){Meech}, {Weryk}, {Micheli}, {Kleyna},
  {Hainaut}, {Jedicke}, {Wainscoat}, {Chambers}, {Keane}, {Petric}, {Denneau},
  {Magnier}, {Berger}, {Huber}, {Flewelling}, {Waters}, {Schunova-Lilly}, \&
  {Chastel}}]{Meech17}
{Meech}, K.~J., {Weryk}, R., {Micheli}, M., {et~al.} 2017, \nat, 552, 378,
  \dodoi{10.1038/nature25020}

\bibitem[{{Mommert} {et~al.}(2019){Mommert}, {Kelley}, {de Val-Borro}, {Li},
  {Guzman}, {Sip{\H{o}}cz}, {{\v{D}}urech}, {Granvik}, {Grundy}, {Moskovitz},
  {Penttil{\"a}}, \& {Samarasinha}}]{Mommert19}
{Mommert}, M., {Kelley}, M., {de Val-Borro}, M., {et~al.} 2019, The Journal of
  Open Source Software, 4, 1426, \dodoi{10.21105/joss.01426}

\bibitem[{{Opitom} {et~al.}(2015){Opitom}, {Jehin}, {Manfroid},
  {Hutsem{\'e}kers}, {Gillon}, \& {Magain}}]{Opitom15}
{Opitom}, C., {Jehin}, E., {Manfroid}, J., {et~al.} 2015, \aap, 584, A121,
  \dodoi{10.1051/0004-6361/201526427}

\bibitem[{{Opitom} {et~al.}(2019){Opitom}, {Fitzsimmons}, {Jehin}, {Moulane},
  {Hainaut}, {Meech}, {Yang}, {Snodgrass}, {Micheli}, {Keane}, {Benkhaldoun},
  \& {Kleyna}}]{Opitom19}
{Opitom}, C., {Fitzsimmons}, A., {Jehin}, E., {et~al.} 2019, \aap, 631, L8,
  \dodoi{10.1051/0004-6361/201936959}

\bibitem[{{Schleicher}(2010)}]{Schleicher10}
{Schleicher}, D.~G. 2010, \aj, 140, 973, \dodoi{10.1088/0004-6256/140/4/973}

\bibitem[{{Yang} {et~al.}(2019){Yang}, {Kelley}, {Meech}, {Keane}, {Protopapa},
  \& {Bus}}]{Yang19}
{Yang}, B., {Kelley}, M. S.~P., {Meech}, K.~J., {et~al.} 2019, arXiv e-prints,
  arXiv:1912.05318.
\newblock \doarXiv{1912.05318}

\bibitem[{{Ye} {et~al.}(2019){Ye}, {Kelley}, {Bolin}, {Bodewits}, {Farnocchia},
  {Masci}, {Meech}, {Micheli}, {Weryk}, {Bellm}, {Christensen}, {Dekany},
  {Delacroix}, {Graham}, {Kulkarni}, {Laher}, {Rusholme}, \& {Smith}}]{Ye19}
{Ye}, Q., {Kelley}, M. S.~P., {Bolin}, B.~T., {et~al.} 2019, arXiv e-prints,
  arXiv:1911.05902.
\newblock \doarXiv{1911.05902}

\end{thebibliography}


\end{document}